\documentclass[preprint,prd,noshowpacs]{revtex4}
\usepackage{amssymb}
\usepackage{amsmath}
\def\be {\begin{equation}}
\def\ee {\end{equation}}
\def\bea {\begin{eqnarray}}
\def\eea {\end{eqnarray}}

\begin{document}
\title{Hawking radiation, chirality, and the principle of effective theory of gravity}

\author{Adamantia Zampeli $^1$} \email[email: ]{a.zampeli@uleth.ca}
\author{Douglas Singleton $^2$} \email[email: ]{dougs@csufresno.edu}
\author{Elias C. Vagenas $^3$} \email[email: ]{evagenas@academyofathens.gr}

\affiliation{$^1$~Theoretical Physics Group, Department of Physics and Astronomy,
University of Lethbridge, 4401 University Drive,
Lethbridge, Alberta, Canada T1K 3M4 \\}

\affiliation{$^2$~Physics Department, CSU Fresno,
Fresno, CA 93740-8031, USA \\}

\affiliation{$^3$~Research Center for Astronomy and Applied
Mathematics, Academy of Athens, Soranou Efessiou 4, GR-11527,
Athens, Greece}

\begin{abstract}
In this paper we combine the chirality of field theories in near horizon regions
with the principle of effective theory of gravity to define a new energy-momentum tensor
for the theory. This new energy-momentum tensor has the correct radiation
flux to account for Hawking radiation for space-times with horizons. This method is
connected to the chiral anomaly cancellation method,
but it works for space-times for which the chiral anomaly cancellation method fails.
In particular the method presented here works for the non-asymptotically
flat de Sitter space-time and its associated Hawking-Gibbons radiation,
as well as Rindler space-time and its associated Unruh radiation. This indicates that it
is the chiral nature of the field theory in the near horizon
regions which is of primary importance rather than the chiral anomaly.
\end{abstract}

\maketitle

%%%%%%%%%%%%%%%%%%%%%%%%%%%%%%%%%%%%
\section{Introduction}
%%%%%%%%%%%%%%%%%%%%%%%%%%%%%%%%%%%%

Hawking radiation from black holes has attracted the interest of theoretical physicists since it was
first discovered by Hawking \cite{Hawking:1974sw}. This physical phenomenon unites gravitational physics near
strong gravitational objects (e.g. black holes) and quantum field theory. Therefore, it can be considered
a milestone of modern theoretical physics and a testbed for candidate theories of Quantum Gravity.
Several methods have been developed for the derivation of Hawking radiation apart from the
seminal one by Hawking \cite{Hawking:1974sw}. Among these methods, the most popular ones are: (i) the tunneling method
\cite{Srinivasan:1998ty,Shankaranarayanan:2000gb,Shankaranarayanan:2000qv,Parikh:1999mf,Vagenas:2001rm},
and (ii) the gravitational chiral anomalies method \cite{Robinson:2005pd,Vagenas:2006qb}. We will discuss a variant
of the latter method here {\it which depends on the chiral nature of field theories in the near-horizon region
but does not depend on the existence of a chiral anomaly}.
We will show it is the chiral nature of the field theory in the near horizon region which
is the key feature rather than the existence of a chiral anomaly since we will examine space-times where the anomaly
vanishes but the near-horizon field theory is still chiral.

The gravitational chiral anomalies method,
in spite of the fact that it has been successfully utilized in many space-times, still faces problems. One
shortcoming with the original chiral anomaly method was that it did not yield the thermal nature of the spectrum of
Hawking radiation. The thermal nature of the spectrum was an assumption in the original work on the anomaly
method. This gap was partially filled by \cite{Iso:2007kt} where higher spin currents were used to calculate
the energy flux and all higher moments of the spectrum. By summing up all the moments one could
obtain a thermal spectrum.  A second problem noted in \cite{Akhmedova:2008au} was that the anomaly method
only worked partially or not at all for the non-asymptotically flat Rindler and de Sitter space-times. In this
paper we will show that it is possible to construct a variant of the gravitational chiral anomaly method
which works for all the space-times for which the original chiral anomaly method works as well
as giving the correct results for those non-asymptotically flat space-times for which
the chiral anomaly method fails. This variant of the anomaly cancellation method relies on the
chirality of the field theory in the near horizon region and the principle of effective theory of gravity
\cite{Padmanabhan:2003ub}, \cite{Padmanabhan:2003gd}.

The paper is organized as follows: (i) we summarize the basic features of the method of gravitational chiral anomalies;
(ii) we construct a new physical approach  based on the property of chirality and the principle of effective theory
of gravity and we discuss the advantages of this new approach; (iii) we derive
the radiation fluxes for Rindler and de Sitter space-times for which the method of gravitational chiral anomalies failed
in part or completely.
\section{Method of gravitational chiral anomalies}
We will illustrate the gravitational chiral anomaly cancellation method for 4-dimensional metrics of the form
\begin{equation}
\label{metric}
ds^2 = -f(r)dt^2 + f(r)^{-1} dr^2 + r^2 d\Omega _{(2)} ~.
\end{equation}
The canonical example is a Schwarzschild black hole which has $f(r)=1-2M/r$ with $M$ the black hole mass.
This metric is asymptotically flat. The location of the event horizon is at $r_H=2M$.
Near the horizon, a quantum field behaves like a $(1+1)$-dimensional free massless scalar field since in this region the kinetic
terms are much larger than the potential terms. Thus the potential terms can be ignored. Fields near the horizon can be analyzed
in terms of an infinite basis of outgoing and ingoing waves. In the Unruh vacuum, the outgoing modes can be identified
as right-moving modes while the ingoing modes as left-moving modes. However, due to the horizon, the ingoing
modes are not accessible to an observer lying outside the black hole horizon and thus these ingoing modes are excluded.
This exclusion results in a chiral theory near the horizon which generates a gravitational chiral anomalies at
the quantum level. These anomalies imply the non-conservation of the energy-momentum tensor $T^\mu_\nu$
in the near-horizon region, although the energy-momentum tensor remains a conserved quantity
far from the horizon. Thus, one splits the space-time outside the horizon into two regions: near-horizon and far-horizon.
The near-horizon physics is described by an energy-momentum tensor which does not satisfy the energy conservation
equation but satisfies an anomaly equation. In the region
far from the horizon the energy-momentum is conserved. The anomaly equation of the consistent
energy-momentum tensor near the horizon, i.e. $T^\mu_{\nu (H)} (r)$, reads \cite{Bertlmann:2000da}
\begin{equation}
\nabla_\mu T^\mu_{\nu (H)} (r)=\pm \mathcal{A}_\nu =\pm \partial_\mu N^\mu_\nu (r)~, ~~~{\rm with}
~~~ N^\mu_\nu (r) = \frac{1}{96\pi}\epsilon^{\beta \mu}\partial_\alpha \Gamma^\alpha_{\nu \beta}~.
\label{consistentanomaly}
\end{equation}
The upper sign is for the case where the accessible modes are the outgoing ones, e.g. when there is an event horizon,
while the lower sign is for the case where the accessible modes are the ingoing ones, e.g. when there is a cosmological horizon.
This correspondence between the upper/lower signs and the accessible outgoing/ingoing modes will be adopted in this paper.
The symbol $\epsilon^{\beta \mu}$ is the totally antisymmetric tensor in two dimensions. The radial
flux of Hawking radiation is defined as
\begin{equation}
\Phi = T^r_{t (H)} (r \rightarrow \infty) ~.
\end{equation}
It is evident that the $\nu =t$ component of (\ref{consistentanomaly}) is needed for the computation of the flux.
For the specific metric given in (\ref{metric}), the consistent anomaly reads \cite{Robinson:2005pd}
\begin{equation}
\mathcal{A}_t (r)= \partial_r N^r_t (r) ~~~~ \mbox{with} ~~~ N^r_t (r)=\frac{1}{192\pi}(f'^{2} + ff'')~.
\label{consistentanomaly1}
\end{equation}
As already mentioned, the energy-momentum tensor far from the horizon, i.e. $T^\mu_{\nu (o)}(r)$, satisfies the conservation equation
\begin{equation}
\nabla_\mu T^\mu_{\nu (o)}(r)=0~.
\label{conservationeq}
\end{equation}
The energy-momentum tensors in (\ref{consistentanomaly}) and (\ref{conservationeq}) are 2-dimensional
and are obtained from the 4-dimensional one, by integrating over the dimensions which are transverse to
the radial dimension, namely
\begin{equation}
T^r_{t (o)} \equiv \int d \Omega_{(2)} r^2 T^r_{t (4)}~.
\end{equation}
After integration with $r_H$  as the lower bound  and $r$ as the upper bound,
the $\nu=t$ component of (\ref{consistentanomaly}) yields
\begin{equation}
T^r_{t (H)}(r)  \mp N^r_t(r) =  T^r_{t (H)}(r_H) \mp N^r_t (r_H)
\label{nearhor}
\end{equation}
while (\ref{conservationeq}) gives $T^r_{t (o)} (r) = c_0 = \Phi $ with $c_0$ being the constant flux at infinity.

The Hawking radiation flux is obtained from the condition that the value of the energy-momentum tensor
far from the horizon should be equal to the limit of the near-horizon energy-momentum tensor at infinity.
For black hole space-times that are  asymptotically flat the quantity $N^r_t (r \rightarrow \infty)$ vanishes.
Therefore, imposing the aforementioned condition, the flux can be written
\begin{equation}
T^r_{t (o)} (r) = \Phi = T^r_{t (H)} (r \rightarrow \infty) = T^r_{t (H)}(r_H) - \frac{1}{192 \pi} f'^2(r_H)~,
\label{conditionasymptflat}
\end{equation}
where we have used (\ref{consistentanomaly1}) to calculate  $N^r_t (r_H)$ and the upper
sign from (\ref{nearhor}) has been used since the radiation is outgoing.
It is now important to introduce the covariant energy momentum and to give its
relation to the consistent energy-momentum tensor \cite{Bertlmann:2000da}
\begin{equation}
\tilde{T}^r_t (r)= T^r_t (r) \pm \frac{1}{192 \pi} (ff''- 2 f'^2)~.
\label{covtensor}
\end{equation}
The covariant condition is that the covariant energy-momentum tensor
should be zero at the horizon, i.e. $\tilde{T}^r_t (r_H)=0$. Using (\ref{covtensor}) and the fact
that $f(r_H) =0$ yields $T^r_t (r_H) = \pm 2\, \frac{f'^2(r_H)}{192 \pi}  = \pm 2 N^{r}_{t}(r_H)$.
By imposing the covariant condition \cite{Banerjee:2007qs}, the flux of Hawking radiation reads
\begin{equation}
\Phi = + N^{r}_{t} (r_H)= \frac{f'^2 (r_H)}{192 \pi}= \frac{\kappa^2}{48\pi} = \frac{\pi}{12} T_H^2 ~,
\label{flux1}
\end{equation}
where $\kappa$ is the surface gravity of the black hole horizon and $T_H$ is the Hawking temperature.
For the metric (\ref{metric}) $\kappa = \frac{f'(r_H)}{2}$ and $T_H = \frac{\kappa}{2\pi}$.

Up to now, we have shown that the flux can be computed when one employs the conservation and anomaly equations
of the consistent energy-momentum tensor. It should be stressed that one can derive exactly the same results
if the conservation and anomaly equations of the covariant energy-momentum tensor are considered.
In particular, the anomaly equation of the covariant energy-momentum tensor reads \cite{Banerjee:2007qs}
\cite{Banerjee:2009} \cite{Majhi:2009}
\begin{equation}
\nabla_\mu \tilde{T}^\mu_\nu (r) =\pm \tilde{\mathcal{A}}_\nu =
\pm \frac{1}{96 \pi \sqrt{-g}}\epsilon_{\nu \mu} \nabla^\mu R ~,
\label{covariantanomaly}
\end{equation}
where $R$ is the Ricci scalar, and $\epsilon_{\mu \nu}$ is the totally antisymmetric tensor in
two dimensions. The $\nu= t$ component of the anomalies is given by
\begin{align}
\tilde{\mathcal{A}}_t = \frac{1}{\sqrt{-g}}\partial_r \tilde{N}^r_t  ~~~
\mbox{with} ~~~
\tilde{N}^r_t (r)= \frac{ff''-\frac{f'^2}{2}}{96\pi}~.
\label{covariantanomaly1}
\end{align}
Following the same analysis and imposing the same boundary conditions as before (i.e.
$\tilde{T}^r_t (r_H)=0$), one finds that the flux is
\begin{equation}
\Phi = -\tilde{N}^r_t (r_H) = \frac{1}{96 \pi} \frac{f'^2 (r_H)}{2} = \frac{\pi}{12} T_H^2 ~.
\label{flux2}
\end{equation}
This flux is identical to one given by the consistent  equations (see (\ref{flux1})),
since  $N^{r}_{t}(r_H) = -\tilde{N}^{r}_{t}(r_H)$.
\section{A new approach based on an effective theory of gravity and chirality}
In this section we introduce a new method for the computation of the radiation flux
from space-times with horizons. The method is related to the gravitational chiral anomaly method
of \cite{Robinson:2005pd,Vagenas:2006qb} but applies to non-asymptotically flat space-times
for which the original method fails. The first observation is that the anomaly method outlined in the previous
section is based on the existence of gravitational chiral anomalies. However, chirality is the consequence of
the exclusion of the modes and precedes the notion of anomalies. There are space-times which are
chiral but for which the covariant anomaly (\ref{covariantanomaly}) vanishes (e.g. de Sitter space-time)
or for which both covariant anomaly (\ref{covariantanomaly}) and consistent anomaly (\ref{consistentanomaly})
vanish (e.g. Rindler space-time). Thus, the critical feature for the new method we propose here
is the existence of chirality of the space-time and not the existence of gravitational chiral anomalies.

The second observation is that we will employ the principle of effective theory of gravity.
The principle of  effective theory of gravity concerns manifolds and observers. It postulates that
{\it ``physical theories in a given coordinate system must be formulated entirely
in terms of the variables that an observer using that coordinate system can access"}
\cite{Padmanabhan:2003ub,Padmanabhan:2003gd}. This principle indicates that one only
needs to take into account those physical quantities/degrees of freedom in the part of
the manifold to which one has access. When calculating things like the thermal Hawking radiation
flux of a given space-time one only needs to consider those degrees of freedom in the part
of the manifold which are on the same side of the horizon as the observer.
One does not have to include degrees of freedom from the part of the manifold
to which the observer does not have access. For a black hole, this means that the physical quantities
that matter are those that exist only on the external manifold. More concretely,
this means that the symmetries we have to take into account are those on the part of
the manifold on which we construct the physical theory.

One should be careful to point out a caveat -- there are cases where some influence or information appears
to leak through the horizon from one region to the other. One example of this is when a pair of quantum
entangled particles is created in a space-time and one of the entangled particles of the pair is allowed
to cross the horizon. This situation was studied in detail in \cite{mann:2005} for the Rindler space-time.
While the presence of the horizon degraded the entanglement there was still some correlation between the
entangled pairs in the presence of the Rindler horizon. Another proposal where some information can
apparently leak across the horizon occurs in the tunneling method of Hawking radiation
\cite{b-zhang:2009, singleton:2010}. In these works it is shown that by taking into account the energy loss
due to Hawking radiation on the mass of the black hole that the spectrum becomes non-thermal and that there
are non-trivial correlations between the emitted Hawking radiation photons. These correlations are shown
to carry away exactly the right amount of information to account for the original Bekenstein entropy of the
black hole. In the present work (as in all other work on the gravitational anomaly method) the spectrum is
assumed to be purely thermal and in this approximation no degrees of freedom on one side of the horizon
can effect degrees of freedom on the other side.

The radial flux from the consistent energy - momentum tensor, i.e. $T^r_t$,
as well as the radial flux from the covariant energy-momentum tensor, i.e. $\tilde{T}^r_t$,
were constructed on the whole manifold  since they were defined under the condition that they are finite on the horizon
with respect to the Unruh vacuum (see \cite{Robinson:2005pd} and \cite{Banerjee:2007qs}, respectively).
According to the method of gravitational chiral anomalies, near the horizon the consistent and covariant
anomaly equations are given by (\ref{consistentanomaly}) and (\ref{covariantanomaly}), respectively.
These equations can be rewritten in the following suggestive form \cite{Papantonopoulos:2008wp}
\begin{equation}
\partial_\mu (T^\mu_\nu \mp N^\mu_\nu) = 0 ~~~
 \mbox{and} ~~~ \partial_\mu (\tilde{T}^\mu_\nu \mp \tilde{N}^\mu_\nu) = 0 ~.
 \label{effectiveconsistent}
\end{equation}
For the $\nu = t$ components this takes the form
\begin{equation}
\partial_r (T^r_t \mp N^r_t) = 0 ~~~
 \mbox{and} ~~~ \partial_r (\tilde{T}^r_t \mp \tilde{N}^r_t) = 0~.
\label{effectivecovariant}
\end{equation}
From the above two sets of equations one can see that the quantities in the parentheses are conserved.
Thus one can use (\ref{effectiveconsistent}) to describe physics both in the
near and far horizon regions, i.e. in the whole part of the manifold external to the horizon.
Utilizing these two observations, one can define a new tensorial quantity as
the {\it effective} consistent energy-momentum tensor that is conserved on the part
of the manifold exterior (the minus sign) or interior (the plus sign) to the horizon
\begin{equation}
{\cal T}^{\mu}_{\nu}= T^{\mu}_{\nu} \mp N^{\mu}_{\nu} ~.
\label{effectiveconsistent1}
\end{equation}
Similarly, one can define an {\it effective} covariant energy-momentum tensor conserved
on the part of the manifold exterior (the minus sign) or interior (the plus sign) to the horizon
\begin{equation}
{\cal \tilde{T}}^{\mu}_{\nu}= \tilde{T}^{\mu}_{\nu} \mp \tilde{N}^{\mu}_{\nu}~.
\label{effectivecovariant1}
\end{equation}
Equations (\ref{effectiveconsistent1}) and (\ref{effectivecovariant1}) are the main results of this work in that
they give an effective energy-momentum tensor which is defined over the entire part of the manifold external/interior to the
horizon rather than being defined piecewise in the near horizon and far horizon regions. These effective
energy-momentum tensors reproduce all the correct results for the Hawking radiation fluxes of asymptotically flat
space-times as the previous gravitational chiral anomaly method, but in addition they will be shown to
work for the non-asymptotically flat de Sitter and Rindler space-times for which the gravitational chiral anomaly method
broke down. In the next two sections, we will calculate
the flux of these space-times employing the effective energy-momentum tensors. This redefinition of the energy-momentum
tensor may be compared to the redefinition of the canonical electromagnetic energy-momentum tensor, $T^{\mu \nu} _{EM}$,
to the symmetric electromagnetic energy-momentum tensor $\Theta ^{\mu \nu} _{EM} = T^{\mu \nu} _{EM} - T^{\mu \nu} _D$,
where $T^{\mu \nu} _D$ is a four-divergence free quantity (see 12.10B of \cite{jackson:1999}). This redefinition of the classical
electromagnetic energy-momentum tensor is required in order satisfy angular momentum conservation, the traceless of the
electromagnetic energy-momentum tensor, gauge invariance, etc. The definitions given above in (\ref{effectiveconsistent1})
and (\ref{effectivecovariant1}) ensure the energy conservation.

Before moving to the calculations of the radiation flux for de Sitter and Rindler space-times we will
generalize (\ref{nearhor}) for our newly defined energy-momentum tensors. Integrating the
first relationship in (\ref{effectivecovariant})
\begin{equation}
 T^r_t (r \rightarrow r_0) \mp N^r_t (r \rightarrow r_0)= T^r_t(r_H) \mp N^r_t (r_H) = \Phi = \pm N^r_t(r_H) ~.
\label{generalconsistent}
\end{equation}
The integration constant, $\Phi$, is the flux defined previously in (\ref{conditionasymptflat}) and (\ref{flux1}).
The upper sign is for an outgoing flux and the lower sign for an ingoing flux.
We have evaluated the integrated quantity at two points $r=r_0$ and $r=r_H$. Previously $r_0 \rightarrow \infty$
since we were dealing with asymptotically flat space-times. We have generalized (\ref{nearhor}) to
(\ref{generalconsistent}) since de Sitter and Rindler space-times are not asymptotically flat so we will need
to take a limit different than $r_0 \rightarrow \infty$. In a similar manner we can integrate the
second relationship in (\ref{effectivecovariant}) to obtain the covariant expression
\begin{equation}
\tilde{T}^r_t (r \rightarrow r_0) \mp \tilde{N}^r_t (r \rightarrow r_0)= \tilde{T}^r_t(r_H) \mp \tilde{N}^r_t (r_H) =
\Phi = \pm N^r_t(r_H)
\label{generalcovariant}
\end{equation}
where the arbitrary point $r_{0}$ is far from all horizons of the space-time where the flux is to be observed
and will be specified from the geometry of the space-time. As previously, the integration constant, $\Phi$,
is the flux defined in (\ref{flux2}) and we have used the fact that $N^{r}_{t}(r_H) = -\tilde{N}^{r}_{t}(r_H)$.
Again  the upper sign is for an outgoing flux and the lower sign for an ingoing flux.
\section{De Sitter spacetime}
The metric of de Sitter space-time in static coordinates is written as
\begin{equation}
ds^2 = -\left( 1-\frac{r^2}{a^2} \right) dt^2 + \left( 1-\frac{r^2}{a^2} \right) ^{-1} dr^2
\end{equation}
with the cosmological horizon located at $r_H=a$. The method of gravitational anomalies can
be applied to de Sitter spacetime only using the consistent expressions of the equations and fails if
one applies the covariant expressions of the equations \cite{Akhmedova:2008au}. The reason for this failure is
that the covariant anomaly vanishes since the Ricci scalar is a constant, i.e. $R=\frac{2}{a^2}$.
However, the elements $N^r_t$ and $\tilde{N}^r_t$ do not vanish at the proper limit, $r_{0}= 0$, namely
$N^r_t(r \rightarrow 0) = \frac{1}{192\pi}\left(-\frac{2}{a^2}\right)$ and
$\tilde{N}^r_t (r \rightarrow 0) =\frac{1}{96\pi}\left(-\frac{2}{a^2}\right)$. The proper limit
is $r_{0}= 0$ since the flux is flowing from the cosmological horizon inwards. The
accessible part of the manifold inside the cosmological horizon and
the flux of Hawking radiation has to be calculated at the point $r_{0}=0$.
Using the equations for the redefined consistent energy-momentum tensor (\ref{effectiveconsistent1})
and (\ref{generalconsistent}) (using the lower signs for these two equations since the flux is coming inward from the
cosmological horizon at $r_H$) we obtain
\begin{equation}
\Phi = {\cal T}^{r}_{t} (r\rightarrow 0) = T^r_t(r_H) + N^r_t (r_H) = - N^r_t(r_H) ~.
\label{flux3}
\end{equation}
Similarly using the equations for the redefined covariant energy-momentum tensor  (\ref{effectivecovariant1})
and (\ref{generalcovariant}) (again using the lower sign in these two equations) we obtain
\begin{equation}
\Phi = {\cal \tilde{T}}^{r}_{t} (r\rightarrow 0) = \tilde{T}^r_t(r_H) + \tilde{N}^r_t (r_H) = - N^r_t(r_H)
\label{flux4}
\end{equation}
It is evident that employing either the redefined effective consistent energy-momentum tensor (\ref{effectiveconsistent1}),
or the redefined effective covariant energy-momentum tensor (\ref{effectivecovariant1}),
the flux of the Hawking radiation in de Sitter space-time is equal to
\begin{equation}
\Phi = - N^r_t (r_H) = +\tilde{N}^r_t (r_H) = - \frac{f'^2 (r_H)}{192\pi} = - \frac{\pi}{12}T_{GH}^2 ~,
\label{flux5}
\end{equation}
where $T_{GH}$ is the Gibbons-Hawking temperature  of the cosmological horizon \cite{Gibbons:1977mu}, i.e.
$T_ {GH}= 1/2\pi a$.
\section{Rindler space-time}
The metric of Rindler space-time in static coordinates reads
\begin{equation}
\label{rindler}
ds^2 = -(1+ 2ar)dt^2 + (1+2ar)^{-1} dr^2 ~,
\end{equation}
where $a$ is the observer's constant acceleration. Due to his acceleration the observer sees a horizon
and hence detects a flux. This is the Unruh effect \cite{Unruh:1976db}. In \cite{Akhmedova:2008au}, it was
shown that both the consistent and covariant anomaly method failed to calculate the Unruh
radiation since the metric in (\ref{rindler}) is simply that of flat space-time as seen by an accelerating
observer and thus both the consistent and covariant anomalies were zero for this space-time. However,
$N^r_t$ and $\tilde{N}^r_t$ where non-zero constant at the limit point $r_0 = 0$, where the static observer resides.
From (\ref{consistentanomaly1}), we obtain $N^r_t = a^2 / 48 \pi$ and from (\ref{covariantanomaly1}), we get $\tilde{N}^r_t =-a^2/48 \pi$
for all $r$. As for de Sitter space-time the limit point is at $r_0 = 0$ rather than infinity. However the horizon
is located at $r_H =-1/2a$ so that the flux is ``outward" from negative $r$ toward positive $r$. Thus in
(\ref{effectiveconsistent1}) and (\ref{effectivecovariant1}) one should upper signs since the flux is ``outward".

The new approach outlined above can be employed here since Rindler space-time is chiral despite the vanishing of both anomalies.
Utilizing the effective consistent energy-momentum tensor (\ref{effectiveconsistent1}) as well
as (\ref{generalconsistent}) (both with the upper sign) the flux reads
\begin{equation}
\Phi = {\cal T}^{r}_{t} (r \rightarrow 0)= T^r_t(r_H) - N^r_t (r_H) = N^r_t(r_H) ~.
\end{equation}
Next using the effective covariant energy-momentum tensor (\ref{effectivecovariant1}) and the
relationship (\ref{generalcovariant}) (again both with the upper sign taken), the flux reads
\begin{equation}
\Phi = {\cal \tilde{T}}^{r}_{t} (r\rightarrow 0) = \tilde{T}^r_t(r_H) -\tilde{N}^r_t (r_H) = - \tilde{N}^r_t(r_H) = N^r_t(r_H) ~.
\end{equation}
In both of the above equations $r_{0} = 0$ and $r_H = -1/2 a$.
It is again evident that employing either the equations of the effective consistent energy-momentum tensor, or
the equations for the effective covariant energy-momentum tensor, the flux of the Unruh radiation in Rindler
space-time is  equal to
\begin{equation}
\Phi = N^r_t (r_H)= - \tilde{N}^r_t (r_H)= \frac{f'^2 (r_H)}{192\pi} = \frac{\pi}{12} T_H^2
\end{equation}
where $T_H= a/ 2\pi$ is the Unruh temperature seen by the accelerated observer.\\
\section{Conclusions}
The new effective energy-momentum tensor, either ${\cal T}^{\mu}_{\nu}$, or ${\cal \tilde{T}}^{\mu}_{\nu}$
helps resolve some of the shortcomings of the gravitational chiral anomaly cancellation
method \cite{Robinson:2005pd, Vagenas:2006qb} for calculating Hawking radiation. Chirality near the horizon
is the symmetry that provides us with the form of the conserved energy-momentum tensor and is responsible
for the existence of the flux. The principle of effective theory of gravity justifies
the extension of the applicability of the equation found near the horizon to the whole effective manifold.
The problems that the definition of this conserved effective energy-momentum tensor resolves,
be it in the consistent or in the covariant form, are:

(a) It provides a physical basis for the calculations of the flux to non-asymptotical space-times
where $N^r_t$ and $\tilde{N}^r_t$ are not zero at infinity while the anomalies,
$\partial_r N^r_t$ and $\partial_r \tilde{N}^r_t$, are zero.
The explanation is that the conserved energy-momentum tensor on the manifold is not  $T^{\mu}_{\nu}$ or $\tilde{T}^{\mu}_{\nu}$
but the  conserved effective energy-momentum tensor ${\cal T}^{\mu}_{\nu}$, or ${\cal \tilde{T}}^{\mu}_{\nu}$, respectively.
Since here we have not considered other sources of energy on the manifold, the constant value of the effective energy-momentum
tensor at infinity should be equal to the flux from the horizon.
However, even in the case of an energy flux from another source, this other source can be inserted in the initial construction
of the effective energy-momentum tensor.

(b) The method applies to Rindler space-time since the theory is still chiral in spite of the fact
that both the consistent and covariant anomalies are zero for the Rindler metric.

(c) The method also applies to de Sitter space-time using the covariant expression of the effective energy-momentum tensor,
while previously only the flux calculation using the consistent expression gave the correct result.

The method outlined here can also be applied with partial success to the time-dependent FRW metric
\begin{equation}
\label{frw}
ds^2= - dt^2 + a^2 (t)\left(\frac{dr^2}{1-kr^2}+r^2d\Omega_{(2)} \right)~.
\end{equation}
Following a series of coordinate transformations \cite{ijmpd:2010}, it is possible to transform the FRW metric
to the de Sitter like form
\begin{equation}
\label{frw1}
ds^2= - (1-\tilde{r}^2/\tilde{r}_A^2)d\tau^2 + \frac{1}{1-\tilde{r}^2/\tilde{r}_A^2}d\tilde{r}^2+\tilde{r}^2d\Omega_{(2)}
\end{equation}
where $\tilde{r}_A=\frac{1}{\sqrt{H^2+k/a^2}}$ and $H = {\dot a}/a$. Applying either the new consistent or the new
covariant energy-momentum tensor method to the FRW metric in form (\ref{frw1}) leads to temperature
$T_{FRW} = 1 / 2 \pi \tilde{r}_A$. This expression is approximately correct since a more complete analysis \cite{sp-kim}
shows that the full expression for $T_{FRW}$ involves an additional terms which depends on ${\dot {\tilde{r}}_A}$.

(d) The method can be generalized in a natural way to space-times with multiple horizons. In these space-times
there are as many energy fluxes as there are horizons. The effective manifold
is that part of the complete manifold between all the horizons. The proper limit for the calculation
is any point in the manifold which is far enough from all the horizons.

(e) It unifies the effective manifold where the energy-momentum tensor is defined
into one region. There is no need to split the manifold into regions near the horizon and
far from the horizon. However, we should be careful about the dimensionality of the theory in each region.
Far from the horizon the theory is n-dimensional while near the horizon we have assumed that the
theory is 2-dimensional. This is a subtle point. The theory is not 2-dimensional.
What we have done is to exclude the transverse dimensions since they play no role in the calculation of
Hawking radiation. We could equally work with the full dimensional theory near the horizon and
then take the Hawking radiation from the (t-r)-dimensions.

\end{document}